\documentclass[aps,prb,twocolumn,showpacs,floatfix]{revtex4}
\usepackage{graphicx}
\usepackage{subfigure}
\usepackage{epsfig}

\begin{document}
\title{Polarization effects in metallic films perforated with a bidimensional array
of subwavelength rectangular holes}
\author{Micha\"{e}l Sarrazin and Jean-Pol Vigneron}
\address{Laboratoire de Physique du Solide\\
Facult\'{e}s Universitaires Notre-Dame de la Paix\\
Rue de Bruxelles 61, B-5000 Namur, Belgium}

\begin{abstract}
For several years, periodical arrays of subwavelength cylindrical holes in
thin metallic layers have taken a crucial importance in the context of the
results reported by Ebbesen {\it et al,} on particularly attractive optical
transmission experiments. It had been underlined that the zeroth order
transmission pattern does not depend on the polarization of the incident
light at normal incidence. In the present paper, we show that it is not the
case for rectangular holes, by contrast to the case of circular holes. In
this context, we suggest a new kind of polarizer that present the advantages
brought by the original Ebbesen devices. Assuming the recent technological
interest for these kinds of metallic gratings, such a kind of polarizer
could lead to new technological applications.
\end{abstract}
\maketitle

\section*{Introduction}

For several years, the motivation for investigating surface plasmon
properties takes a crucial importance in the context of plasmonics. In
addition, the consideration of metallic gratings have been renewed in the
context of the results report by Ebbesen {\it et al }[1]{\it .} They
reported on particularly attractive optical transmission experiments on
periodical arrays of subwavelength cylindrical holes in a thin metallic
layer deposited on glass, a dielectric substrate. Two specific
characteristics of their results are the transmission which is higher than
the simple sum of individual holes contributions and the interesting pattern
of the zeroth order transmission versus wavelength [1].

During the last few years, these experimental results have received
considerable attention and they have implied several theoretical and
experimental works [1-17]. Many assumptions were suggested to explain the
transmission properties of these devices. For instance, it has been
suggested that these phenomena can be described in terms of short range
diffraction of evanescent waves [11], or in terms of dynamical diffraction
[13]. Another explanation suggest the role of cavity resonance into the
holes to explain the transmission enhancement [12,15].

However, the main results tend to prove that surface plasmons (SPs) play a
key role in the features observed in the transmission curves [1-7,9,10].
Nevertheless, it remains many questions which must be answered to clarify
the processes involved in these experiments. For instance, the exact role of
SPs was not clearly assessed. Indeed, the observed transmission presents a
set of peaks and minima. Many authors suggested that transmission peaks
correspond to SPs resonance [1-7,9,10] whereas others suggested that SPs
resonance were related to minima [12,14].

In a recent work [14], we have presented many results which tend to qualify
these two hypothesies. We have shown that the concept of resonant Wood
anomalies [14,18,19] can be invoked to interpret the role of SPs in the
Ebbesen experiments. We have shown that the transmission pattern is better
described by Fano's profiles [19] correlated with interferences between non
resonant processes and the resonant response of SPs coupled with
nonhomogeneous diffraction orders [14]. We have shown that each maximum of
transmission (preceded by a minimum) corresponds to one maximum of a Fano's
profile (preceded by the related minimum) [14,19]. Moreover, whereas such
Fano's profile in transmission is related to a resonant process, the
location of its maximum (or minimum) does not necessarily correspond to the
location of the resonance [14,19].

In such devices made of subwavelength hole arrays, polarization properties
start only to be underlined in some experimental works [16,17]. In the
original experiments, for square gratings made of circular holes in chromium
films for instance [4], it has been underlined that the zeroth order
transmission pattern does not depend on the polarization of the incident
light at normal incidence [4]. Obviously, for a square grating made of
rectangular holes for instance, one can expect a polarization dependent
transmission. Indeed, for a square grating with circular holes the main
axies of the grating can be inverted whereas for a square grating with
rectangular holes these axies cannot be inverted without break the grating
symmetry.

In the present paper, we will show the polarization dependency of the
transmission for rectangular holes arrays contrary to circular holes arrays.
We will present a simple analytical interpretation of our numerical results.
However, we will underline the limits of this simple interpretation as one
must not forgot the real complexity of the problem which yet requires a
numerical study. Despite a simple appearance, the present problem is not
trivial. Our results will lead us to suggest a new kind of polarizer which
presents the advantages brought by the original Ebbesen devices [1,4].

\section*{Numerical approach and studied device}

In the following, our simulations rest on a coupled modes method associated
with the use of the scattering matrix ({\bf S} matrix) formalism. Taking
into account the periodicity of the device, the permittivity is first
described by a Fourier series. Then, the electromagnetic field is described
by Bloch's waves which can also be described by a Fourier series. In this
context, Maxwell's equations take the form of a matricial first order
differential equation along the $z$ axis perpendicular to the $x$ and $y$
axies where the permittivity is periodic [14,20,21]. The heart of the method
is to solve this equation. One approach deals with the propagation of the
solution step by step by using the {\bf S} matrix formalism. More
explicitly, we numerically divide the grating along the $z$ axis into many
thick layers for which we calculate the {\bf S} matrix. The whole {\bf S}
matrix of the system is obtained by using a special combination law applied
step by step to each {\bf S} matrices along the $z$ axis. Indeed, it is well
know that {\bf S} matrices and their combinations are much better
conditionned than transfert matrices [21]. In this way, we can calculate the
amplitudes of the reflected and transmitted fields, for each diffracted
order (which corresponds to a vector ${\bf g}$ of the reciprocal lattice)
according to their polarization ($s$ or $p$) [14]. Note that our algorithm
has been accurately compared with other methods such as FDTD or KKR [22]. In
the present work the convergence is obtained from two harmonics only, i.e.
for 25 vectors of the reciprocal lattice. Furthermore, there is no
convergence problem associated with discontinuities here, and we do not need
to use Li's method [23,24].

As a continuation of our previous works, we turn our attention to the case
of chromium films [14]. The values of chromium and glass permittivities are
those obtained from experiments [25]. For instance, glass permittivity is
approximately equal to $2.10$ for wavelengths ranging from $1000$ to $1500$ $%
nm$. Here, the thickness of the metallic film is taking equal to $h=100$ $nm$%
. The film is perforated with rectangular holes of sides $a=350$ $nm$ and $%
b=700$ $nm$ respectively. The shorter side is along the $x$ axis and the
longer side is along the $y$ axis. The holes shape a square grating of
parameter $c=1000$ $nm$ (fig.1a). Incident light is normal to the interface
i.e. normal to the $Oxy$ plane. The incident light is linearly polarized and
the orientation of the incident electric field ${\bf E}_{inc}$ is given by
the angle $\theta $ between ${\bf E}_{inc}$ and the $y$ axis (fig.1a) in the
$Oxy$ plane. Because the transmitted zeroth order is normal to the $Oxy$
plane, its polarization is described in the $Oxy$ plane too. The transmitted
zeroth order electric field will be described by the complex electric field
components $E_x$ and $E_y$ along the $x$ and $y$ axies respectively. The
polarization of the transmitted zeroth order is then described by the
modulus of the amplitudes $\left| E_x\right| $ and $\left| E_y\right| $ and
the dephasing $\delta =\arg \left\{ E_y/E_x\right\} .$ The amplitude of the
incident light is equal to $1$ $V.m^{-1}$.

\section*{Results and analysis}

First, we represent in fig.1b a schematic view of the reciprocal lattice
related to the square grating of paramater $c$ in reduced coordinates. Each
point corresponds to a reciprocal lattice node, i.e. to one diffraction
order ${\bf g}=\frac{2\pi }c(i,j)$. Solid lines represent a contours
representation of the modulus of the Fourier transform of the rectangular
hole. For the most common diffraction orders we give the values of the
modulus of the Fourier transform corresponding to the rectangular hole. Such
values are representative of the typical order of magnitude for the
relatives amplitudes of the diffraction orders. For instance the orders $%
(\pm 1,0)$ are associated with a relative amplitude of $0.810$ whereas it is
of $0.318$ for the orders $(0,\pm 1)$, i.e. $2.5$ times less. Obviously, for
a circular hole, the Fourier transform would be symmetric if we substitute $%
i $ by $j$ and vice versa. Then, the relative amplitude of the orders $(\pm
1,0)$ and $(0,\pm 1)$ would be equal. This difference between circular and
rectangular holes is crucial. Indeed, in our recent work [14] we have
specified the role of SPs in the Ebbesen experiment through the concept of
resonant Wood anomalies [14,18,19]. As explained above, the transmission
pattern is described by Fano's profiles correlated with interferences
between non resonant processes and the resonant response of the surface
plasmons coupled with nonhomogeneous diffraction orders. More precisely, the
key process in those device appears as follows. We suppose that the incident
light is polarized along the $[1,0]$ axis of the grating. The incident light
then diffracts against the grating and generates nonhomogeneous resonant
diffraction orders e.g. $\left( 1,0\right) $ which is $p$ polarized. Such a
resonant order is coupled with a surface plasmon (obviously $\left(
0,1\right) $ is also generated but it is $s$ polarized and then it is not a
resonant order, i.e. it is not coupled with surface plasmons). It becomes
possible to excite this surface plasmon which leads to a feedback reaction
on the order $\left( 1,0\right) $. Then this diffraction order\ can diffract
too and generate a contribution to the homogenous zeroth diffraction order $%
\left( 0,0\right) $. In fact $s$ polarized diffraction orders give no
contribution to the enhancement of the transmission, contrary to $p$
polarized diffraction orders [14]. Obviously, the contribution of the order $%
\left( 1,0\right) $ depends on the relative amplitude related to the Fourier
transform of the hole. Now, if the incident light is polarized along the $%
[0,1]$ axis of the grating, the order $\left( 0,1\right) $ becomes the
resonant $p$ polarized order whereas $\left( 1,0\right) $ becomes the
non-resonant $s$ polarized order. Then two cases occur. If the Fourier
transform is symmetric, the substitution of the $p$ polarized order $\left(
1,0\right) $ by the $p$ polarized order $\left( 0,1\right) $ let invariant
the transmission. If the Fourier transform is not symmetric, then the
contribution to zeroth order of both $p$ polarized orders will be not the
same according to the incident polarization. In our case, if the incident
light is polarized along the $[1,0]$ axis of the grating, the $p$ polarized
order $\left( 1,0\right) $ (which is associated with a relative amplitude of
$0.810$ of the Fourier transform) gives an important contribution to the
zeroth order transmission. If the incident light is polarized along to the $%
[0,1]$ axis of the grating, the $p$ polarized order $\left( 0,1\right) $
(which is associated with a relative amplitude of $0.318$ of the Fourier
transform) gives a weak contribution to the zeroth order transmission. Then
the transmission $t_x$ for light polarized along the $x$ axis will be
greater than the transmission $t_y$ for light polarized along the $y$ axis.
This let to some issues. As explained above, the electric field of the
incident light can be written as
\begin{equation}
{\bf E}_{inc}=E_{inc}\left[
\begin{array}{c}
\sin \theta \\
\cos \theta
\end{array}
\right]
\end{equation}
Assumming the fact that both components of the incident electric field are
not transmitted in the same way, the zeroth order transmitted field is
written as
\begin{equation}
{\bf E}=\left[
\begin{array}{c}
E_x=t_xE_{inc}\sin \theta \\
E_y=t_yE_{inc}\cos \theta
\end{array}
\right]
\end{equation}
where $t_x$ and $t_y$ respectively are the transmission coefficients along
the $x$ and $y$ axis respectively. The modulus of the amplitudes are then
\begin{equation}
\left\{
\begin{array}{c}
\left| E_x\right| =\left| t_xE_{inc}\right| \sin \theta \\
\left| E_y\right| =\left| t_yE_{inc}\right| \cos \theta
\end{array}
\right.
\end{equation}
As the zeroth order transmission can be written as
\begin{equation}
T=\sqrt{\frac{\varepsilon _s}{\varepsilon _v}}\frac 1{\left| E_{inc}\right|
^2}\left( \left| E_x\right| ^2+\left| E_y\right| ^2\right)
\end{equation}
it is easier to show that one can write $T$ as
\begin{equation}
T=T_{90^{\text{o}}}+\left( T_{0^{\text{o}}}-T_{90^{\text{o}}}\right) \cos
^2\theta
\end{equation}
where $T_{0^{\text{o}}}=\sqrt{\frac{\varepsilon _s}{\varepsilon _v}}\left|
t_y\right| ^2$ and $T_{90^{\text{o}}}=\sqrt{\frac{\varepsilon _s}{%
\varepsilon _v}}\left| t_x\right| ^2$ ($\varepsilon _s$ and $\varepsilon _v$
are the permittivities of the substrate and the vaccum respectively). At
last, note that the dephasing $\delta $ is such that
\begin{equation}
\delta =\arg \left\{ \frac{t_y}{t_x}\right\}
\end{equation}
One shows that eq.5 is similar to the well known Malus Law for polarizer. To
illustrate our assumptions, fig.2 shows the zeroth order transmission
against wavelength for different $\theta $ values. We clearly show the
dependence of the transmission pattern on the incident polarization. For $%
\theta =90^{\text{o}},$ i.e. the incident electric field is along the $x$
axis, the transmission is maximal and identical to what one observes in the
Ebbesen experiments [1,4]. It is shown that the transmission increases with
the wavelength, and that it is characterized by sudden changes in the
transmission marked 1 to 2 on the figure. These values are shifted toward
larger wavelengths when the grating size increases [14]. Moreover, for $%
\theta =0^{\text{o}}$, i.e. when the incident electric field is along the $y$
axis, the transmission is minimal. This result is, at least qualitatively,
in agreement with our hypothesis. Note that wavelengths 1 and 2 are close to
Rayleigh's wavelengths. In diffraction gratings such wavelengths are those
for which a diffracted order becomes tangent to the plane of the grating. At
normal incidence, and for a diffraction order ${\bf g}=\frac{2\pi }c(i,j)$,
Rayleigh's wavelength is defined as
\begin{equation}
\lambda _R^{u,i,j}=c\sqrt{\varepsilon _u}(i^2+j^2)^{-\frac 12}
\end{equation}
where $\varepsilon _u$\ represents either the permittivity of the vacuum ($%
\varepsilon _v$), or of the dielectric substrate ($\varepsilon _d$). For the
diffraction orders such that $(\pm 1,0)$ or $(0,\pm 1),$ the values of
Rayleigh's wavelengths in the present case are $1000$ $nm$ for the
vacuum/metal interface, and $1445.29$ $nm$ for the substrate/metal interface.

Fig.3 shows two zoomed plots around wavelengths 1 (fig.3a) and 2 (fig.3b).
In fig.3a, one shows that the transmission minima are localized at the same
wavelength whatever $\theta .$ Surprisingly, in this narrow wavelength
domain, transmission for $\theta =0^{\text{o}}$ is greater than for $\theta
=90^{\text{o}}$ contrary to previously. The minimum which corresponds to the
wavelength 1, is located at $1001.44$ $nm$ just after the related Rayleigh's
wavelength. Fig.3b shows that transmission minima are weakly shifted toward
large wavelengths as $\theta $ increases (minima locations for each $\theta $
value, are marked by vertical black dashs). For $\theta =0^{\text{o}}$, the
minimum is located at $1445.29$ $nm$, i.e. the wavelength 2 which
corresponds to the related Rayleigh's wavelength. For $\theta $ greater than
zero, minimum is shifted. Moreover, one notes a sudden change in the
transmission at the Rayleigh's wavelength whatever $\theta $. The minima
observed here correspond to minima of Fano's profiles [14]. In fig.3a, the
minima are not of the same kind of the minima in Fig.3b. This is not true
minima of the Fano's profile. All occur as if the minimum of the Fano's
profile disappears behind the Rayleigh's wavelength towards low wavelength.
In other words, the minima in Fig.3a come from the cutoff and the
discontinuity introduced in Fano's profiles at the Rayleigh's wavelength.
This had been widely explained in ref. [14].

For more details, in fig.4, we have reported the wavelengths $\lambda _m$
related to the transmission minima (related to fig.3b) against $\theta $
(black dots)$.$ For comparison, and by analogy with eq.5, the solid line is
related to the following equation
\begin{equation}
\lambda _{m,\theta }=\lambda _{m,90^{\text{o}}}+(\lambda _{m,0^{\text{o}%
}}-\lambda _{m,90^{\text{o}}})\cos ^2\theta
\end{equation}

Quantitatively, fig.5 shows the variation of the transmission against the
angle of polarization $\theta $ for both wavelengths $1300$ $nm$ and $1900$ $%
nm$ (fig.5a) and both wavelengths $1147.64$ $nm$ and $1001.44$ $nm$ (fig.5b)
in the neighborhood of the above mentionned minima. The numerical
computations give results indicated by the dots on the figure. Using the
numerical values of the transmission for $\theta =90^{\text{o}}$ and $\theta
=0^{\text{o}}$ respectively, i.e. $T_{90^{\text{o}}}$ and $T_{0^{\text{o}}}$%
, one plots the transmission against $\theta $ for both wavelength by using
eq.5 (solid lines). One observes then a quantitative agreement between the
pattern from eq.5 and the results obtained numerically. Note the weakly
decreasing transmission at $1001.44$ $nm$ as already mentionned, which is
also in full agreement with eq.5.

In order to more detail our approach, we present on fig.6a the amplitudes $%
\left| E_x\right| $ and $\left| E_y\right| $ of the zeroth transmitted order
against wavelength for two different values of $\theta .$ One shows that the
$\left| E_y\right| $ amplitude does not exhibit high convex regions in
transmission contrary to the $\left| E_x\right| $ amplitude. As explained,
the convex regions in the zeroth order transmission (i.e. the high
transmission domains between minima $1$ and $2,$ and just after minimum $2$)
are induced by the scattering of non-homogenous resonant diffraction orders
such as $(\pm 1,0)$ or $(0,\pm 1)$. For the $\left| E_x\right| $ amplitude,
the orders $(\pm 1,0)$ are associated with a relative amplitude of $0.810$
of the Fourier transform of the hole profile. This large contribution of
these orders enables an enhanced transmission via multi-scattering. This
effect appears as the reason of the existence of the convex region of high
transmission as shown in previous work [14]. On the contrary, for the $%
\left| E_y\right| $ amplitude the resonant orders $(0,\pm 1)$ are associated
with a relative amplitude of $0.318$ of the Fourier transform of the hole
profile. This contribution is not large enough to obtain a significant
enhanced transmission. For $\theta =30^{\text{o}}$, the $\left| E_y\right| $
amplitude contribution to the whole transmission is greater than the $\left|
E_x\right| $ contribution. Although the transmission along the $x$ axis is
greater than along the $y$ axis, as ${\bf E}_{inc}$ components depend on $%
\theta $ (see eq.1) the $x$ axis component of the incident electric field is
small enough to lead $\left| E_x\right| $ to be lower than $\left|
E_y\right| $ (see eq.2). On the contrary for $\theta =60^{\text{o}}$, the $%
\left| E_x\right| $ amplitude contribution to the whole transmission is
greater than the $\left| E_y\right| $ contribution. Indeed, the transmission
along the $x$ axis is obviously greater than along the $y$ axis, but now the
$x$ axis component of the incident electric field is also greater than the
one along the $y$ axis. Then, the amplitude $\left| {\bf E}\right| $ pattern
for $\theta =60^{\text{o}}$ is essentially defined by the pattern of the
amplitude $\left| E_x\right| $ whereas the amplitude $\left| {\bf E}\right| $
pattern for $\theta =30^{\text{o}}$ is essentially defined by the pattern of
the amplitude $\left| E_y\right| $. In addition, the whole zeroth order
transmission would be greater for $\theta =60^{\text{o}}$ than for $\theta
=30^{\text{o}}.$ Again, these results are in qualitative agreement with our
assumptions. Quantitatively, fig.6b shows the variation of the amplitudes $%
\left| E_x\right| $ and $\left| E_y\right| $ against the angle of
polarization $\theta $ for both wavelength $1300$ $nm$ and $1900$ $nm$. The
numerical computations give the results indicated by the dots on the figure.
Using the numerical values of the amplitudes for $\theta =90^{\text{o}}$ and
$\theta =0^{\text{o}}$, one plot the transmission against $\theta $ for both
wavelength by using eq.3 (solid lines). As previously, one observes a
quantitative agreement between the pattern from eq.3 and the results
obtained numerically.

In the possibility of later experimental results, we wish to add some
details for comparison. Fig.7a gives the dephasing $\delta $ against
wavelength for both angle $\theta =60^{\text{o}}$ and $\theta =30^{\text{o}%
}. $ One shows the peaks close to the minima of transmission, i.e. close to
the surface plasmon resonance according to our previous work [14]. Despite
the absence of an explicit dependence against $\theta $ for $\delta $
according to eq.6, one notes that such a dependence exist after all. So,
fig.7b shows the dephasing against $\theta $ for both wavelength $1300$ $nm$
and $1900$ $nm$. One shows for $\theta =0$ that the dephasing is equal to
zero, and the polarization of the zeroth transmitted order is thus the same
as the incident field. Though the dephasing tends to be different from zero
when $\theta =90^{\text{o}}$, as $\left| E_y\right| $ tends to zero (see
fig.6) the polarization is the same as the incident field. By contrast, the
transmission polarization will be elliptical for others values of $\theta $.
Such a dependence of $\delta $ against $\theta $ and wavelength probably
express a dependence of $t_x$ and $t_y$ against $\theta $. This maybe
corresponds to a propagating constant against $z$-axis that is different
according to the $x$-axis or $y$-axis. In addition, as shown in ref.[14], $%
t_x$ and $t_y$ result from many contributions such as SPs via
multi-scattering. Because SPs exhibit a polarization dependency, this could
explain the $t_x$ and $t_y$ behaviour. Nevertheless, as one shows, the
dependence of $\delta $ against $\theta $ and wavelength is not well
understood and we hope to clarify this point in a later work. Moreover, it
will be necessary to search for the effects induced by a parameters change.

At last, it is striking that our numerical results coincide with our simple
and intuitive explanation which leads to predict that a square grating with
rectangular holes can behave like a polarizer. Indeed, the whole
multiscattering processes are complex and there are numerous contributions
to the zeroth order transmission as explained above. It was then not
necessarily obvious, that our intuitive model match up with our numerical
results, all the more that the results shown in fig.7 underline the limits
of our simple interpretation.

\section*{Conclusion}

In the present paper, we have shown that contrary to the original
experiments (which used a grating made of circular holes) [4], the zeroth
order transmission pattern depends on the polarization of the incident light
when using rectangular holes. The results presently obtained match up with
our previous work [14]. In this context, we obtain a device with properties
globally similar to those of the original Ebbesen experiments, but which
exibits in addition, some properties of a polarizer. Assuming the recent
technological interest for this kind of metallic gratings, we think that the
kind of device introduced in the present paper could lead to new
technological applications. At last, note that, during the submission
process, Gordon et al [17] have published experimental results concerning
transmission through a bidimensional array of elliptical holes in metallic
film. Their results clearly establish the polarization dependence of the
transmission with asymmetric holes. These experimental results corroborate
the present theoretical approach and in this way, the experimental and
theoretical works supply many complementary results.

\section*{Acknowledgments}

We acknowledge the use of Namur Scientific Computing Facility (Namur-SCF), a
common project between the FNRS, IBM Belgium, and the Facult\'{e}s
Universitaires Notre-Dame de la Paix (FUNDP).

This work was carried out with support from EU5 Centre of Excellence
ICAI-CT-2000-70029 and from the Inter-University Attraction Pole (IUAP P5/1)
on ''Quantum-size effects in nanostructured materials'' of the Belgian
Office for Scientific, Technical, and Cultural Affairs.

\section*{References}

[1] T.W. Ebbesen, H.J. Lezec, H.F. Ghaemi, T. Thio, P.A. Wolff, Nature
(London) 391, 667 (1998)

[2] H.F. Ghaemi, T. Thio, D.E. Grupp, T.W. Ebbesen, H.J. Lezec, Phys. Rev.
B, 58, 6779 (1998)

[3] U. Schr\"{o}ter, D. Heitmann, Phys. Rev. B, 58, 15419 (1998)

[4] T. Thio, H.F. Ghaemi, H.J. Lezec, P.A. Wolff, T.W. Ebbesen, JOSA B, 16,
1743 (1999)

[5] J.A. Porto, F.J. Garcia-Vidal, J.B. Pendry, Phys. Rev. Lett., 83, 2845
(1999)

[6] D.E. Grupp, H.J. Lezec, T.W. Ebbesen, K.M. Pellerin, T. Thio, Appl.
Phys. Lett. 77 (11) 1569 (2000)

[7] T. Thio, H.J. Lezec, T.W. Ebbesen, Physica B 279, 90 (2000)

[8] E. Popov, M. Nevi\`{e}re, S. Enoch, and R. Reinisch, Phys. Rev. B 62,
16100-16108 (2000)

[9] A. Krishnan, T. Thio, T. J. Kim, H. J. Lezec, T. W. Ebbesen, P.A. Wolff,
J.\ Pendry, L. Martin-Moreno, F. J. Garcia-Vidal, Opt. Commun., 200, 1 (2001)

[10] L. Martin-Moreno, F.J. Garcia-Vidal, H.J. Lezec, K.M. Pellerin, T.
Thio, J.B. Pendry, T.W. Ebbesen, Phys. Rev. Lett., 86, 1114 (2001)

[11] J.-M. Vigoureux, Optics Comm., 198, 4-6, 257 (2001)

[12] Q. Cao, Ph. Lalanne, Phys. Rev. Lett., 88 (5), 057403 (2002)

[13] M.M.J. Treacy, Phys. Rev. B 66, 195105, (2002)

[14] M. Sarrazin, J.-P. Vigneron, J.-M. Vigoureux, Phys. Rev. B, 67, 085415
(2003)

[15] Z. Sun, Y.S. Jung, H.K. Kim, Appl. Phys. Lett. 83 (15) 3021 (2003)

[16] E. Altewischer, M.P. van Exter, J.P. Woerdman, JOSA B, 20, 1927 (2003)

[17] R. Gordon, A. G. Brolo, A. McKinnon, A. Rajora, B. Leathem, and K. L.
Kavanagh, Phys. Rev. Lett. 92, 037401 (2004)

[18] R.W. Wood, Phys. Rev. 48, 928 (1935)

[19] V.U. Fano, Ann. Phys. 32, 393 (1938)

[20] J.P. Vigneron, F. Forati, D. Andr\'{e}, A. Castiaux, I. Derycke, A.
Dereux, Ultramicroscopy, 61, 21 (1995)

[21] J.B. Pendry, P.M. Bell, NATO\ ASI\ Series E Vol. 315 (1995)

[22] V. Lousse, K. Ohtaka, Private Communication (2001)

[23] L. Li, JOSA\ A 13\ (9) 1870 (1996)

[24] L. Li, JOSA A 14 2758 (1997)

[25] D.W. Lynch, W.R. Hunter, in Handbook of Optical Constants of Solids II,
edited by E.D. Palik (Academic Press, New York, 1991)

\section*{Captions}

Figure 1 : (a) Diagrammatic view of the system unders study, (b) Simplified
view of the Fourier transform of the holes grating.

Figure 2 : zeroth order transmission against wavelength for various
polarization, i.e. for various values of $\theta .$

Figure 3 : close view of zeroth order transmission against wavelength for
various polarization, i.e. for various values of $\theta .$ (a) in the
vicinity of wavelength (1). (b) in the vicinity of wavelength (2).

Figure 4 : Wavelengths related to the transmission minima against $\theta $
in the vicinity of wavelength (2) (see fig.3b).

Figure 5 : (a), zeroth order transmission against $\theta $ for both
wavelength $1300$ $nm$ and $1900$ $nm$. (b), zeroth order transmission
against $\theta $ for both wavelength $1001.44$ $nm$ and $1447.64$ $nm$.

Figure 6 : (a) $\left| E_x\right| $ and $\left| E_y\right| $ amplitudes
against wavelength for $\theta =30^{\circ }$ and $\theta =60^{\circ }.$ (b) $%
\left| E_x\right| $ and $\left| E_y\right| $ amplitudes against $\theta $
for both wavelength $1300$ $nm$ and $1900$ $nm$.

Figure 7 : (a) dephasing $\delta $ against wavelength for $\theta =30^{\circ
}$ and $\theta =60^{\circ }.$ (b) dephasing $\delta $ against $\theta $ for
both wavelength $1300$ $nm$ and $1900$ $nm$.

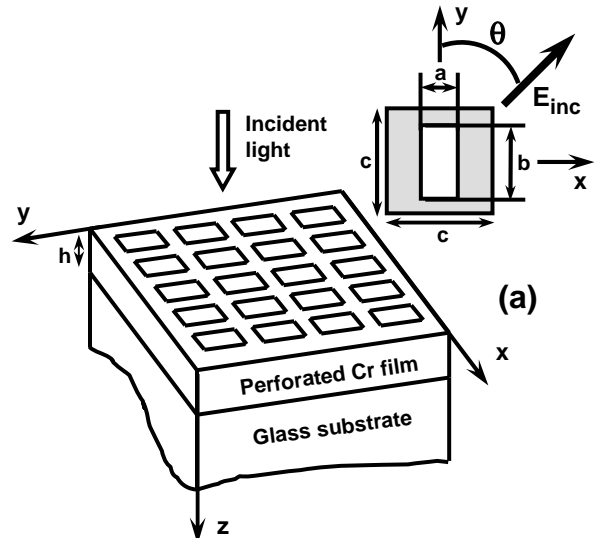
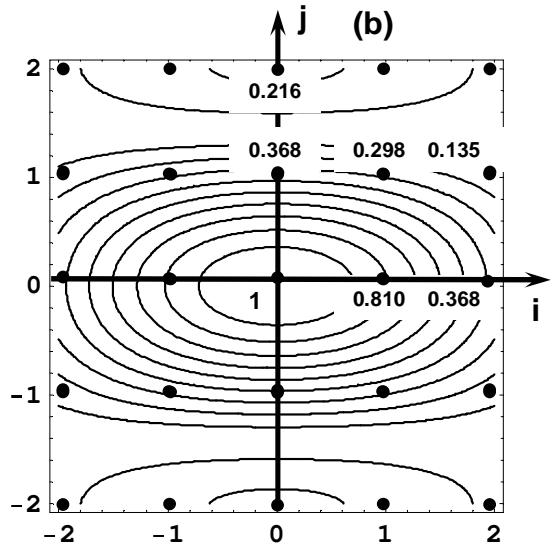
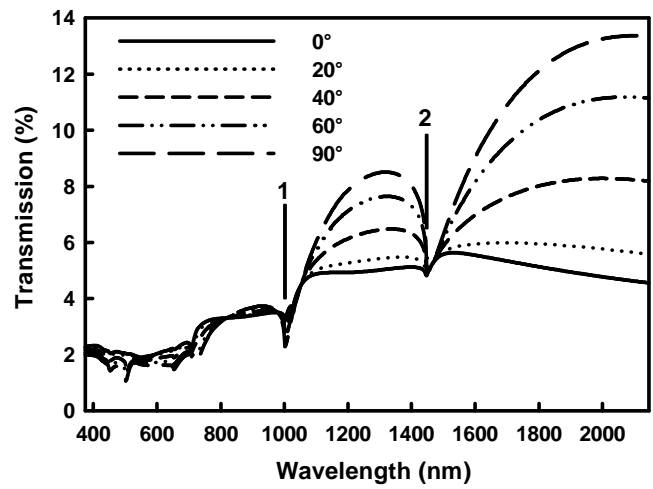
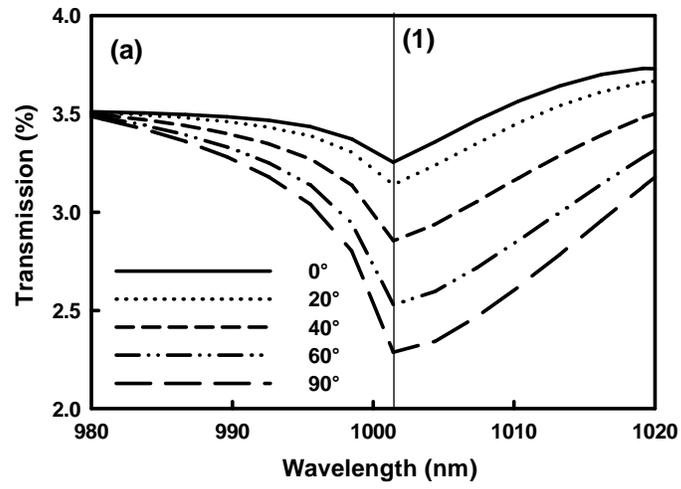
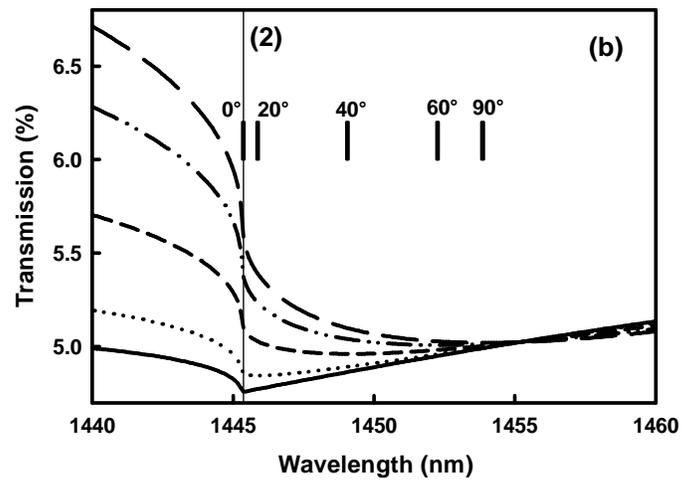
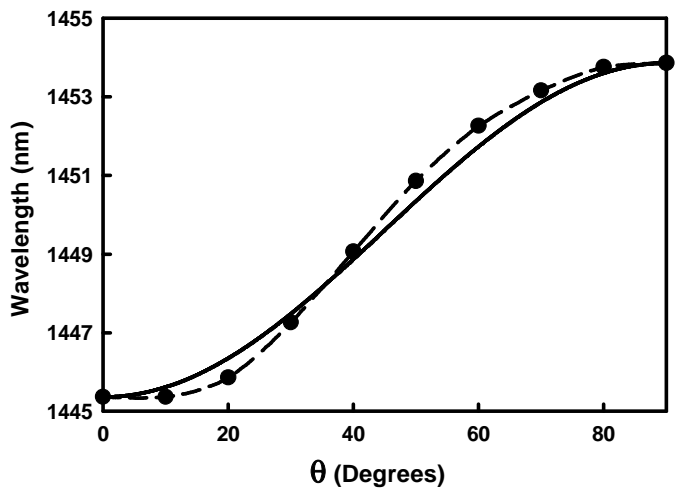
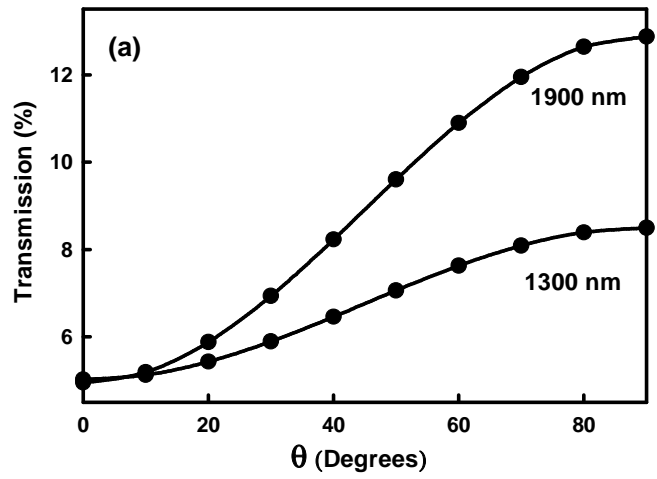
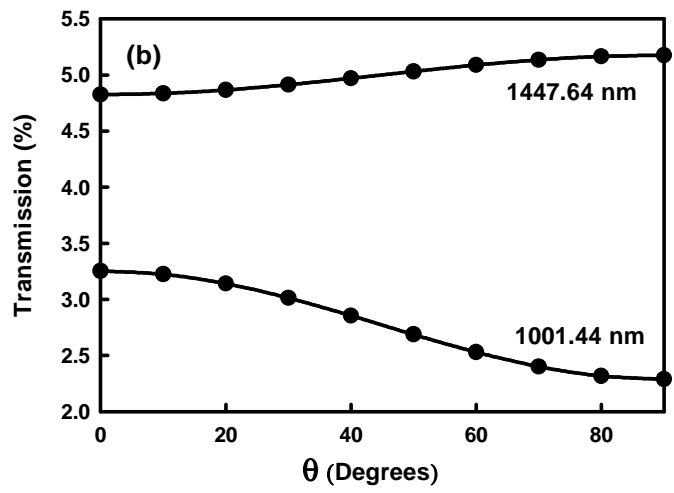
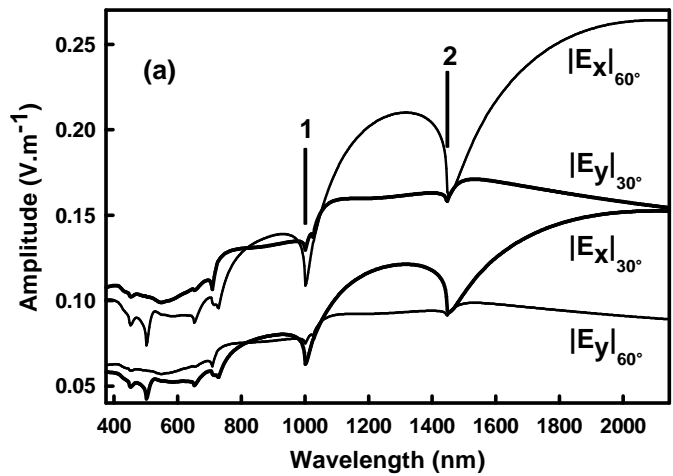
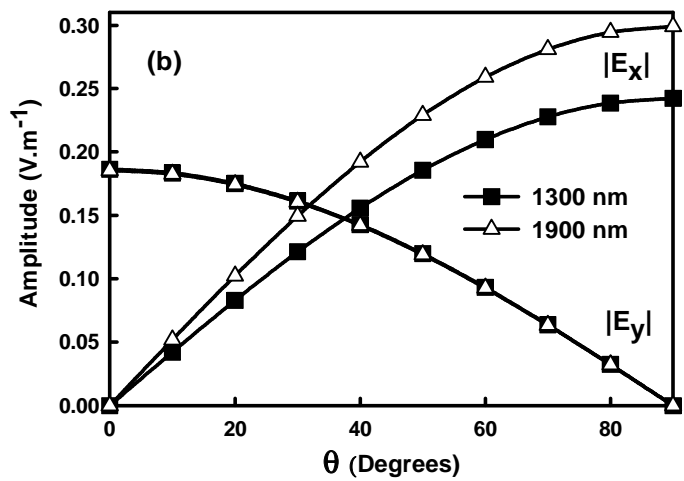
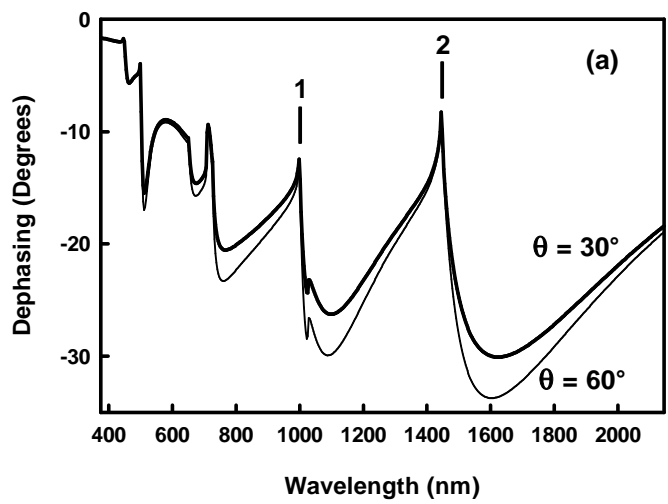
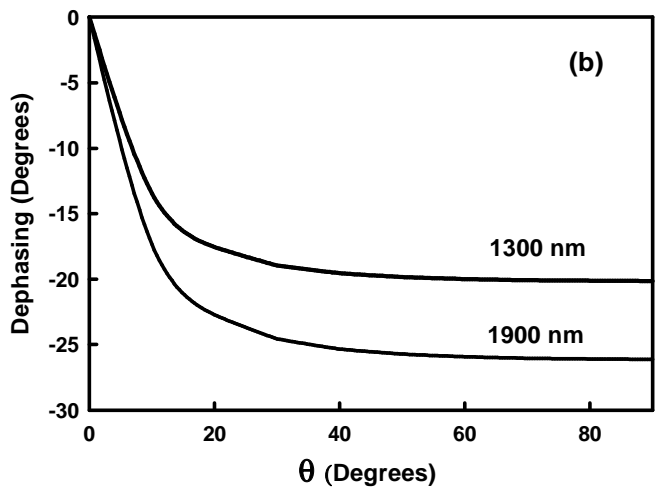
\end{document}